\documentclass[preprint,showpacs,preprintnumbers,amsmath,amssymb]{revtex4-1}

\usepackage{graphicx,epsfig}
\usepackage{dcolumn}
\usepackage{bm}
\newcommand{\be}{\begin{equation}}
\newcommand{\ee}{\end{equation}}
\newcommand{\bse}{\begin{subequations}}
\newcommand{\ese}{\end{subequations}}
\newcommand{\bary}{\begin{eqnarray}}
\newcommand{\eary}{\end{eqnarray}}
\newcommand{\bwt}{\begin{widetext}}
\newcommand{\ewt}{\end{widetext}}

\begin{document}

\title{High energy neutrinos from choked GRBs and their flavor ratio
  measurement by the IceCube. }
\author{
 Karla Varela$^{1,3}$, Sarira Sahu$^2$, Andr\'es Felipe Osorio
 Oliveros$^1$, 
 Juan Carlos Sanabria$^1$
}
\affiliation{$^1$Universidad de Los  Andes, Bogota, Colombia\\
  $^2$Instituto de Ciencias Nucleares, Universidad Nacional Aut\'onoma de M\'exico, \\
  Circuito Exterior, C.U., A. Postal 70-543, 04510 Mexico DF, Mexico\\
 $^3$Max-Planck-Institut fur Extraterrestrische Physik, Giessenbachstrasse 1, 85748, Garching, Germany
}

\begin{abstract}

The high energy neutrinos produced in a choked GRB can undergo matter
oscillation before emerging out of the stellar envelope. Before
reaching the detector on Earth, these neutrinos can undergo further vacuum
oscillation and then Earth matter oscillation. In the context of
IceCube we study the Earth matter effect on neutrino flux in the
detector.  For the calculation of track-to-shower ratio R in the
IceCube, we have included the shadowing effect and the additional contribution from the muon track produced by
the high energy tau lepton decay in the vicinity of the detector.
We observed that R is different for different CP phases in
vacuum but the matter effect suppresses these differences. We have also
studied the behavior of R when the spectral index $\alpha$ varies.


\end{abstract}
\maketitle

\section{Introduction}
\label{intro}

Gamma-Ray Bursts (GRBs) are cosmological events with the emission of
very intense electromagnetic radiation in the energy range $\sim$ 100 keV - 1 MeV.  
Phenomenologically GRBs come in two
variants: the short-hard bursts and long-soft bursts. The 
long gamma-ray bursts (LGRBs, typically with duration longer than 
2 seconds), which constitute about 3/4 of the total observed GRBs,
are generally believed to be associated with deaths of massive 
stars\cite{Zhang:2003uk,Kouveliotou:1993yx}. In this scenario the gamma rays emitted by the collapsing star during
 a long GRB event should be the result of relativistic jets of
 radiation and matter breaking through the stellar
 envelope. Fermi-accelerated electrons would produce gamma rays by
 synchrotron and inverse Compton scattering in optically thin
 magnetized relativistic shocks. In this same shock protons should
 also be accelerated to relativistic velocities and interact with the
 photons producing neutrinos with an energy range from  MeV-
 EeV\cite{Waxman:1997ti,Waxman:1998yy}. Observationally, only a small fraction ($\leq 10^{-3}$) of core
collapse SNe are associated with GRBs\cite{Berger:2003kg,Mazzali:2003np,Woosley:2006fn}.
These correspond to the cases when the energetic jet successfully 
penetrates through the stellar envelope and reaches a highly 
relativistic speed (Lorentz factor $\Gamma \geq 100$). It is possible
that  the larger fraction of the core collapse may not be able to
punch through the massive envelope to launch a successful
GRB. Irrespective of its failure to emerge out from the thick envelop, like the successful jet, these choked jet can also
accelerate protons to very high energy and produce multi-TeV neutrinos
through interaction with  the keV photon background present
in the jet environment\cite{Sahu:2010ap}. The high energy neutrinos are
produced from the decay of charged
pions which lead to the neutrino flux ratio at the source
$\Phi^0_{\nu_e}:\Phi^0_{\nu_\mu}:\Phi^0_{\nu_\tau}=1:2:0$
($\Phi^0_{\nu_{\alpha}}$ corresponds to the sum of neutrino and
antineutrino flux at the source). As is well known, matter effect can
substantially modify the flux ratio due to neutrino oscillation, in
a presupernova star scenario, high energy neutrinos propagating
through  a heavy envelope can oscillate to other flavors due to  matter effects,
resulting in flavor ratios at the surface of the star that can be significantly
different from 1:2:0. In a
previous paper\cite{Oliveros:2013apa} 
(Paper-I) we presented a detail calculation of the effects of matter inside the
presupernova star on the neutrino fluxes, using a 
formalism that takes into account the three neutrino flavors and
different density profiles for the presupernova star. 
Our results show that for neutrinos with $E_{\nu} \leq 10$ TeV the
fluxes on the surface of the star are different from the original one
1:2:0. We have also calculated the fluxes of the these neutrinos on the
surface of the Earth after they travel through the long baseline
between the source and the Earth. We found that for neutrino energy $E_{\nu} \le
10$ TeV, the flux ratio is different from 1:1:1 and above this energy
the ratio converges to 1:1:1 implying that matter effect does not play
a significant role for high energy neutrinos.

The IceCube neutrino detector in South pole is fully operational since
December 2010. The IceCube collaboration has reported the observation of 37 neutrino events in the
energy range 30 TeV-2 PeV and the sources of these events are unknown\cite{Aartsen:2013bka,Aartsen:2013jdh,Aartsen:2014gkd}.
These neutrino events have flavors, directions and energies
not compatible with the atmospheric neutrinos and it is believed that
this is the first indication of extraterrestrial origin of high
energy neutrinos. Recently, IceCube collaboration has presented results of 641
days data taken during 2010-2012  in the energy range 1 TeV-1
PeV from the southern sky which gives a new constraint on the diffuse
astrophysical neutrino spectrum\cite{Aartsen:2014muf}. These high energy neutrino events
have generated much interest and several models are proposed for
their origin. The choked GRBs are potential candidates to produce the
high energy neutrinos which can propagate hundreds of Mpc baseline to
reach the Earth. So it is important to study these neutrinos and the
matter effect on their propagation. The present work is an extension of Paper-I. Here we
take into account the matter effect of both presupernova star medium and
the Earth on the calculation of the flux ratio by a detector like
IceCube which could be relavent to get information regarding the type
of progenitor responsible for the choked GRBs. We also take into
account the shadowing effect of Earth on these neutrinos.

The organization of the papers is as follows:
In Sec.2 we discuss about the neutrino propagation in the Earth by
considering the realistic density profile of it. Here we also take
into account the shadowing effect which is important for high energy
neutrinos. In Sec. 3, the signature of shower and track events are
discussed. The detailed calculation of track-to-shower ration is
discussed in Sec. 4. Finally we present our results in Sec. 5 followed
by a summary in Sec. 6.

\section{Matter effect on neutrinos going through the Earth}
\label{sec:1}


The energy spectra of the gamma rays produced by long GRBs have been
measured and they  follow power laws, or broken power 
laws\cite{Halzen:1999xc}. In the GRB jet (both successful and choked),
neutrinos are produced with varying energy depending on the distance
from the central engine. The one which are closer to the central
engine are in the MeV range and it increases as the distance
increases. This happens because the protons are Fermi accelerated
within the jet  and gain energy as the distance increases up to a
maximum, where neutrinos of $\sim$ EeV energy can be produced.
In this environment the high energy $\gamma$-rays and neutrinos are
produced through $pp$ and/or $p\gamma$ interaction within the jet
environment and the fluxes of
these GeV-TeV neutrinos and the $\gamma$-rays are related. Both the
$\gamma$-rays and the neutrinos have power-law spectrum. Here we assume a
simple power-law spectrum for the high energy neutrinos as:
\be
 \frac{dF_{\nu_l}}{dE_{\nu_l}} = N_{\nu_l} {E_{\nu_l}}^{-\alpha},
 \label{eq:Nu_Espectrum}
\ee
where $\alpha\ge 2$ is the spectral index and $N_{\nu_l}$ is the normalization
constant in units of $GeV^{-1} cm^{-2} s^{-1}$.

\begin{figure}[!ht]
\begin{center}
\includegraphics[width=90mm, height=60mm]{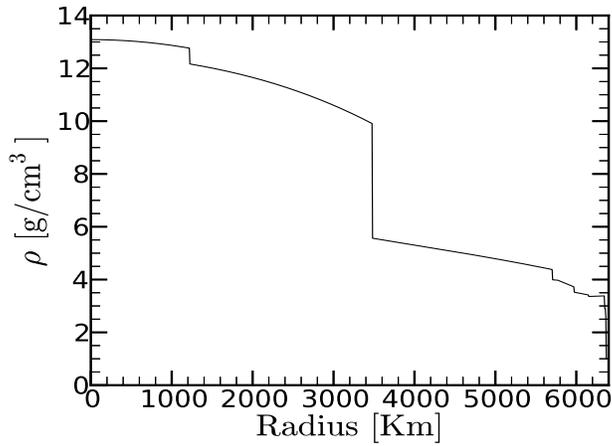} 
\caption{Earth matter density profile $\rho_e$\ \  [g/cm$^3$] as a
  function of  radius.}
\label{fig:earthmatter}
\end{center}
\end{figure}

High energy neutrinos reaching the detector on Earth from the opposite side
can experience absorption due to neutrino-nucleon CC and NC interactions. For very high energy
neutrinos the interaction cross sections are large enough so that the
absorption effects become very important and have to be taken into
account. The shadowing factor due to this absorption is 
given by\cite{Becker:2007sv}:
\be
 P_{\text{shad}} = \text{exp}(- N_{A} \sigma_{TOT} X),
\label{eq:Pshadow}
\ee
where $\sigma_{TOT}$ is the total neutrino-nucleon  cross section,
$N_{A} = 6.0221 \times 10^{23}$ mol$^{-1}$ is the Avogadro's number,
and $X$ is the column depth traveled by the neutrino inside the Earth
before interaction. The column depth is the product of the distance traveled and the
density of matter inside the Earth $\rho_e$.  Since the Earth's density
depends on position, $\rho_e = \rho_e(r)$ and $\ X$ is given by:
\be
 X = \int{\rho_e(r)\  d\mathbf{r}}\ ,
\label{eq:Column}
\ee
where the integral is a path integral along the trajectory of the
neutrino, from the entrance point to the Earth up to the detector, and
can be parametrized in terms of the zenith angle $\theta$ of the
neutrino track at the detector. The cross section $\sigma_{TOT}$ is a
function of the neutrino energy $E_{\nu}$. Then the shadowing factor
$P_{shad}$ depends on both $E_{\nu}$ and $\theta$ and can be expressed
as $P_{shad} =P_{shad}(E_{\nu},\theta)$. We consider the most realistic density
profile of the Earth, which is given by\cite{Becker:2007sv}:
\be
 \rho_{e} \left(r\right) = \left\{ 
  \begin{array}{l l}
    13.0885 - 8.8381\cdot x^{2} & \quad x < 0.192\\
    12.5815-1.2638 \cdot x\  - & \\
     \ \ 3.6426 \cdot x^{2} - 5.5281 \cdot x^{3} & \quad 0.192 < x < 0.546 \\
    7.9565-6.4761 \cdot x\ + & \\ 
    \ \ 5.5283 \cdot x^{2} - 3.0807 \cdot x^{3} & \quad 0.546 < x < 0.895 \\
    5.3197-1.4836 \cdot x & \quad 0.895 < x < 0.906 \\
    11.2494-8.0298 \cdot x & \quad 0.906 < x < 0.937 \\
    7.089-3.8045 \cdot x & \quad 0.937 < x < 0.965 \\
    2.691+0.6924 \cdot x & \quad0.965 < x < 0.996 \\ 
    2.9 & \quad 0.996 < x < 0.998\\
    2.6 & \quad 0.998 < x < 0.999\\
    1.02 & \quad 0.999 < x \leq 1, \\
  \end{array} \right.
  \label{eq:earthprofile}
\ee
where  $x = r/R_{Earth}$ and $\rho_e$ is given in units of g/cm$^3$. The Earth
density profile is shown in Fig. \ref{fig:earthmatter}. Using this
density profile $X(\theta)$ can be calculated.

\begin{figure}[!ht]
\begin{center}
\includegraphics[width=90mm, height=60mm]{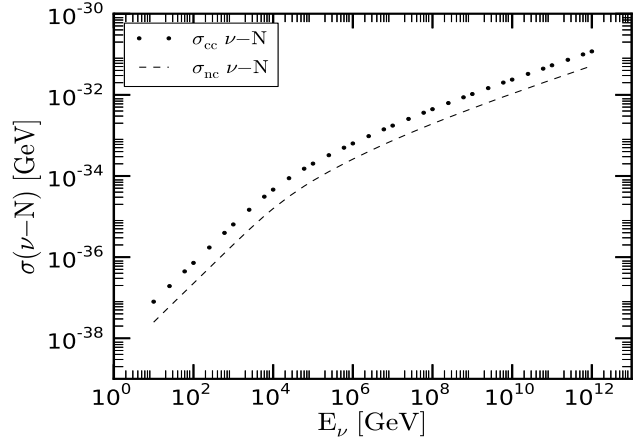} 
\caption{Neutrino-nucleon cross sections at high energies \cite{Gandhi:1998ri}.}
\label{fig:Xsec_neutrinos}
\end{center}
\end{figure}

\begin{figure}[!ht]
\begin{center}
\includegraphics[width=90mm, height=60mm]{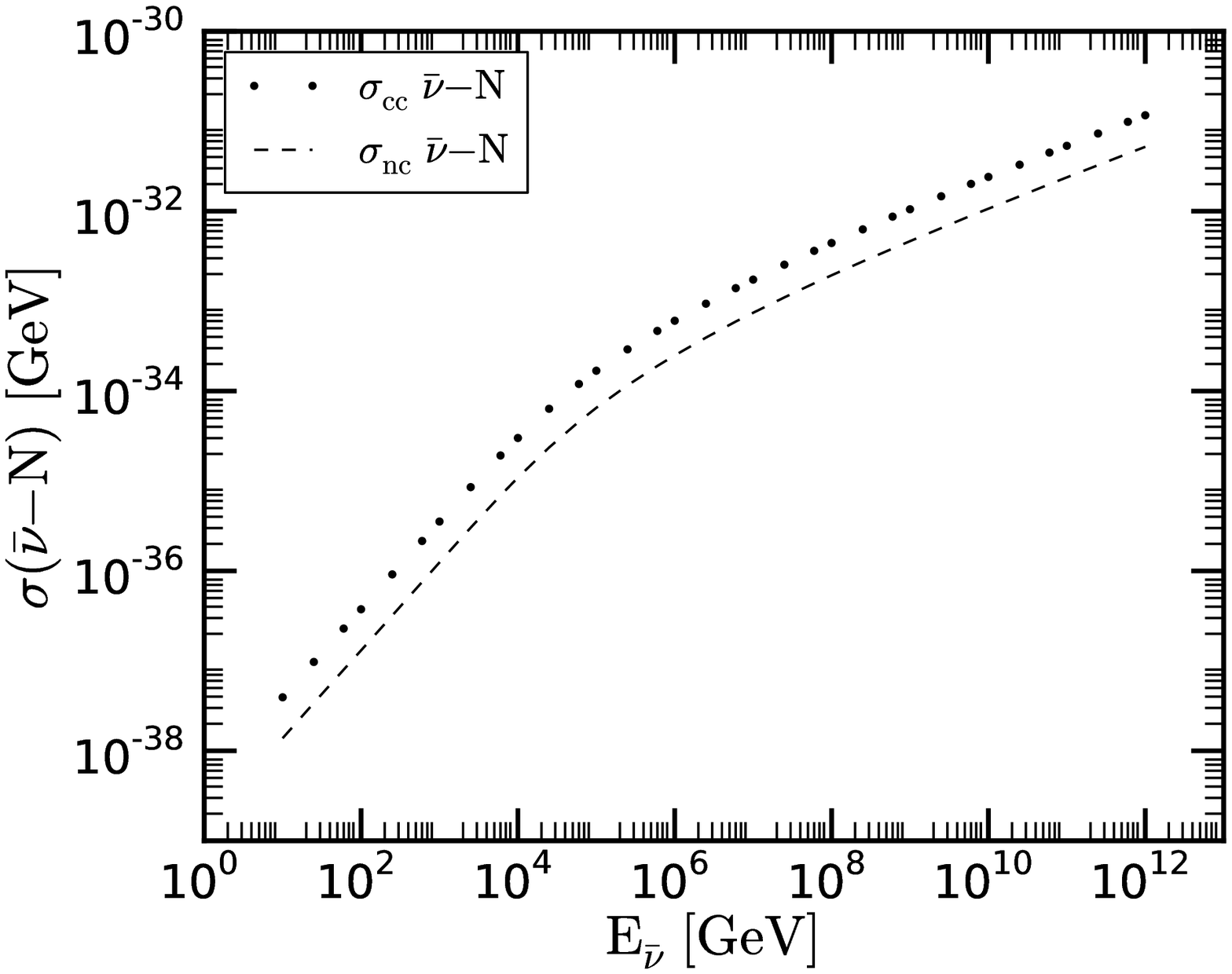} 
\caption{Antineutrino-nucleon cross sections ant high energies \cite{Gandhi:1998ri}.}
\label{fig:Xsec_antineutrinos}
\end{center}
\end{figure}

\begin{figure}[!ht]
\begin{center}
\includegraphics[width=90mm, height=60mm]{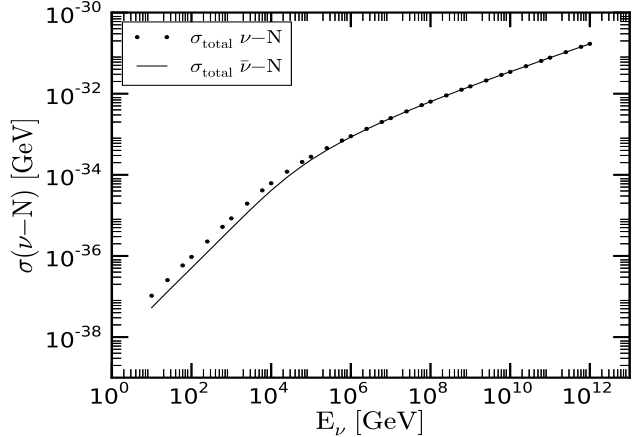} 
\caption{Comparison of both neutrino-nucleon and antineutrino-nucleon cross sections.}
\label{fig:Xsec_total}
\end{center}
\end{figure}

The values of the total cross sections, for neutrino and antineutrino
interaction with matter (nuclei) at high energies, have to be extrapolated from
low energy data, since no measurements  have been performed yet. In this
work we use the cross sections reported in Ref.\cite{Gandhi:1998ri}
and present in Figs. \ref{fig:Xsec_neutrinos} and
\ref{fig:Xsec_antineutrinos} respectively for $\nu- N$ and ${\bar \nu}
-N$.  Comparison of the total cross sections $\nu- N$ and ${\bar \nu}
-N$ shows that in the low energy limit $E_{\nu} \le 10$ TeV
there is a very small difference between these two which can be seen
in Fig. \ref{fig:Xsec_total}.

\begin{figure}[!ht]
\begin{center}
\includegraphics[width=80mm, height=60mm]{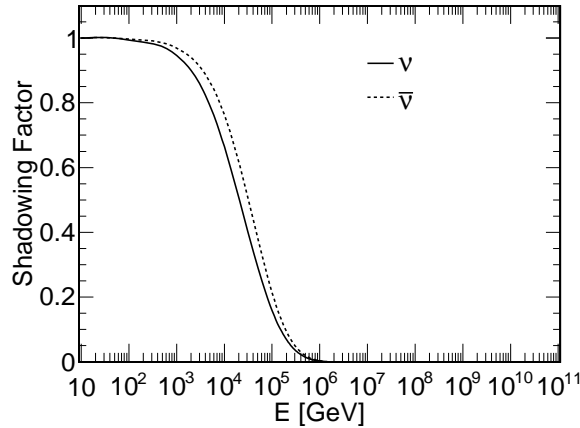} 
\caption{The shadowing factor $P_{shad}$ as a function of neutrino energy for a zenith angle $\theta = 180^{\circ}$.}
\label{fig:shad}
\end{center}
\end{figure}

In Fig. \ref{fig:shad},\  $P_{shad}$ is plotted as a function of
$E_{\nu}$, for a zenith angle  $\theta = 180^{\circ}$ (neutrinos
arriving to the detector from underneath).  From the graph it can be
noticed that the shadowing factor decreases as the neutrino energy
increases beyond $\sim 1$ TeV and the Earth becomes opaque for neutrinos with
energies above $\sim 1000$ TeV. There is a small difference between
neutrino and antineutrino shadowing factor above 1 TeV.  Since we are interested in TeV
neutrinos, the shadowing effect has to be taken into account properly
in the calculation of neutrino fluxes arriving at the detector.
Depending on the energy of the neutrinos, the interaction of the neutrinos with the medium inside the Earth will
also result on flavor oscillations.  Since in this work we will
account for those neutrinos that go through the Earth before undergoing deep inelastic collision with the surround
medium to the detector, we must take into account the flavor
oscillation.

In Paper-I we have already used the analytic
formalism developed by T. Ohlsson and H. Snellman (OS) to calculate
three-flavor neutrino oscillations\cite{Ohlsson:1999xb,Ohlsson:2001et} in the
presupernova star\cite{Oliveros:2013apa} and then calculate  the flavor
ratio of neutrinos arriving on Earth. Here we are extending the
calculation by taking into account
the matter effect of the Earth to calculate the flavor ratio at the
IceCube detector. For this
calculation we use the Earth density profile given in Eq. (\ref{eq:earthprofile}).

\begin{figure}
\begin{center}
\includegraphics[width= 80mm, height=80mm]{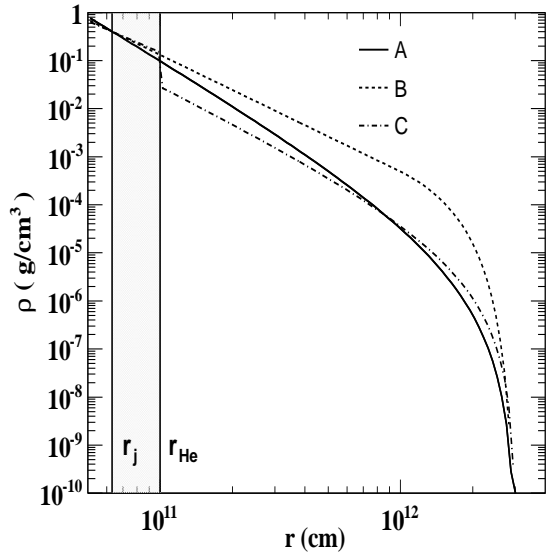}   
\caption{Density profiles of the progenitor star taken from
  \cite{Mena:2006eq}. The density profiles [A], [B], [C] are described in
  detail in \cite{Oliveros:2013apa}.}
\label{fig:profiles}
\end{center}
\end{figure}

\begin{figure}
\begin{center}
\includegraphics[width= 80mm, height=80mm]{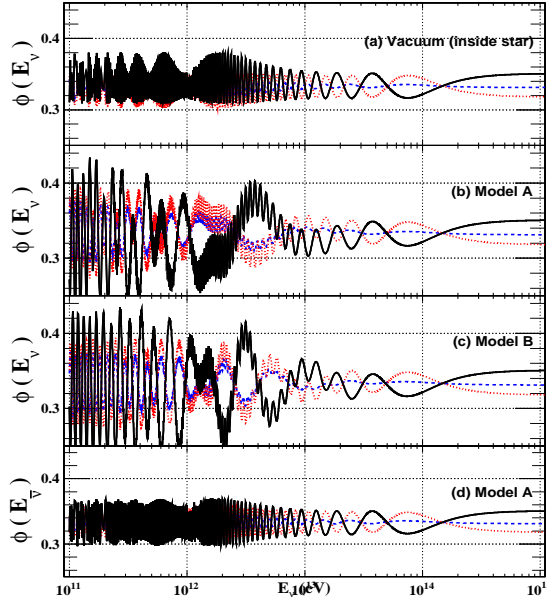}   
\caption{Neutrino and antineutrino fluxes at the detector. In
  (\textbf{a}), (\textbf{b}) and (\textbf{c}) \emph{solid lines} is
  for $\Phi_{\nu_e}$, \emph{dashed lines} are for $\Phi_{\nu_{\mu}}$,
  and \emph{dotted lines} are for $\Phi_{\nu_{\tau}}$. In (\textbf{d})
  \emph{solid line, dashed line} and \emph{dotted line} are for
  $\Phi_{\bar{\nu}_{e}}$,  $\Phi_{\bar{\nu}_{\mu}}$ and
  $\Phi_{\bar{\nu}_{\tau}}$ respectively. The neutrino-mixing
  parameters are: $\delta_{CP} = 0$; $\theta_{12} = 33.8^{\circ}$;
  $\theta_{13} = 8.8^{\circ}$; $\theta_{23} = 45.0^{\circ}$; $\Delta m
  ^{2}_{21} = 8.0 \times 10^{-5}$ eV$^2$; $\Delta m ^{2}_{23} = 3.2
  \times 10^{-3}$ eV$^2$.} 

\label{fig:Neut_FluX_S1}
\end{center}
\end{figure}

\begin{figure}
\begin{center}
\includegraphics[width= 80mm, height=80mm]{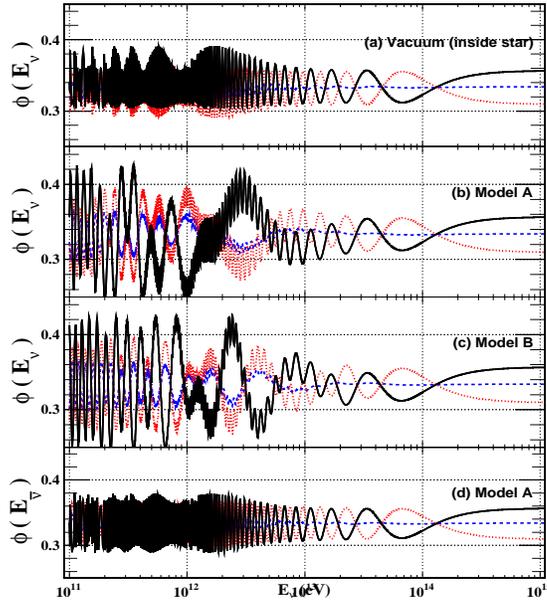}   
\caption{Same as Fig. \ref{fig:Neut_FluX_S1} but here we consider $\theta_{13} =
  12.0^{\circ}$. }
\label{fig:Neut_FluX_S2}
\end{center}
\end{figure}

The input neutrino fluxes at the surface of the Earth, as functions of
neutrino energy $E_{\nu}$, are those calculated in Paper-I, for three
different models of the presupernova star, which we will refer to
as model A, B and C and are discussed throughly in Paper-I. For
reference we present the density profile of these three models in
Fig. \ref{fig:profiles} and a detail description is given in paper-I\cite{Oliveros:2013apa}.
In Figs. \ref{fig:Neut_FluX_S1} and \ref{fig:Neut_FluX_S2} the
neutrino and antineutrino
fluxes at the detector, as functions of neutrino energy, resulting from
the models  A and B  (in (b), (c) and (d)) and taking into account the
Earth's matter effect, are compared with the case in
which the effects of the stellar medium are ignored (in (a)).
The two sets of plots, corresponding to different  neutrino-mixing angles
$\theta_{13}$, are shown. In these plots the neutrinos have traversed the whole Earth before arriving to the
detector (a 180$^{\circ}$ zenith angle). All other parameters are taken from the best
fit parameters from different experiments which are surrarized in
Table \ref{tab1}. We also consider two sets of parameters Set-I and Set-II corresponding to two different
presuprenova star radii $R^*$ as shown in Table \ref{tab1} and analyze
our results.

\begin{table}
\centering
\caption{We consider these parameters for our study. $R^*$ is the
  radius of the presupernova star. We consider three different values of
$\Delta m^2_{32}$ to observe the variation in $R$. }
\label{tab1}
\begin{tabular*}{\columnwidth}{@{\extracolsep{\fill}}llll@{}}
\hline
\multicolumn{1}{@{}l}{Parameter} &Set-I &Set-II\\
\hline
$R^{*}$  & $3\times 10^{12}\,cm$  & $2.7\times 10^{12}\, cm$\\
$\theta_{12}$  & $33.8^{\circ}$  & $33.8^{\circ}$\\
$\theta_{13}$  & $8.8^{\circ}$   & $12^{\circ}$\\
$\theta_{23}$  & $45^{\circ}$    & $45^{\circ}$\\
$\Delta m^2_{21}/eV^2$  &  $8.5\times 10^{-5}$  &  $8.5\times 10^{-5}$\\
\hline
$\Delta m^2_{32}/eV^2$  & $1.4\times 10^{-3}$  & $1.4\times 10^{-3}$\\
\,\, & $3.2\times 10^{-3}$  &  $3.2\times 10^{-3}$\\
\,\, & $6.0\times 10^{-3}$  &  $6.0\times 10^{-3}$\\
\hline
\end{tabular*}
\end{table}

\section{Detection of neutrinos by IceCube}
\label{sec:2}


A neutrino detector, like IceCube, detect high energy neutrinos by
observing the Cherenkov radiation emitted by the secondary charged particles
produced when high energy neutrinos interact with the
surrounding rock and ice\cite{icecubesite}. These secondaries produce showers events and/or
tracks events depending on the primary neutrino flavor. The neutrino
interaction with rock and ice takes place through
neutral current (NC) and/or charge current 
(CC) weak processes
$\nu_{l}+N\rightarrow \nu_{l} ({l})+ X$.  In the NC case, since there is a neutrino in the
final state, the only signature of the interaction will be through the
hadronic shower, independent of the neutrino
flavor. In the CC case the end-result
depends on neutrino flavor. If the interacting neutrino is an
electron type, the resulting electron will quickly interact with the medium,
producing an electromagnetic shower, which will overlap with the
hadronic shower. If the neutrino is  muon type, the resulting muon will produce a
long track that emerges from the shower. Finally, if the  neutrino is tau
type, the resulting tau lepton may or may not produce a track
depending on its energy.
But when the tau
decays into muon, $\tau\rightarrow \nu_{\mu} \, \mu\, \nu_{\tau}$ 
the later will produce a long track, just like in the case of a muon-neutrino CC
interaction, this modifies the number of track events, which has to be
accounted for. Since in this
work we are considering neutrinos coming from underneath the detector,
those with energies above 1 PeV will be drastically suppressed, and
therefore the {\it lollipop}  and {\it double-bang} events that are associated with
very energetic $\nu_{\tau}$ will also be suppressed\cite{Beacom:2003nh}. In this work we
will not consider these kinds of events, however, we will include the 
$\mu$-track events induced by tau neutrinos, as explained above.

In conclusion, the ratio of track events to shower events is related
in a convoluted way to the neutrino  flavor
ratios. However, given a set of flavor ratios, like 1:1:1 in the
"standard picture", or any other set, like in the case we are
presenting in this work, the ratio of tracks-to-showers R can be
calculated. In the next section
we discuss in detail the track-to-shower ratio calculation.

\begin{figure}
\begin{center}
\includegraphics[width= 80mm, height=80mm]{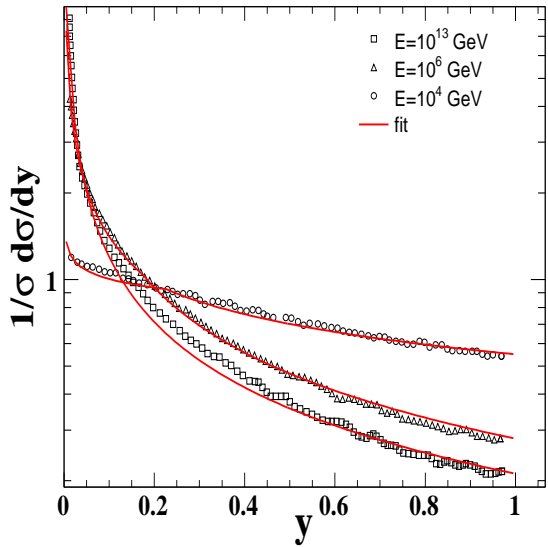}   
\caption{Comparison between our empirical fit, based on equation
  (\ref{eq:dsdy}), 
and the Monte Carlo results presented in reference \cite{Bevan.:2007pt}.}
\label{fig:dsdy}
\end{center}
\end{figure}

\section{The track-to-shower ratio}
\label{sec:3}


The calculation of the track-to-shower ratio R presented in this
section is based on the calculations from 
references\cite{Beacom:2003nh, Esmaili:2009dz}. Here we have included the
shadowing effect due to the neutrino absorption by the Earth,
$P_{shad}(E_{\nu},\theta)$. Since we are considering neutrinos coming
from underneath, $\theta = 180^{o}$, then $P_{shad} (E_{\nu})=
P_{shad}(E_{\nu},\theta=180^{\text{o}})$. The ratio R is defined as:
\be
\text{R} = \frac{\text{Number of}\  \mu\text{-track events}}{\text{Number of shower-like events}}.
\label{eq:R}
\ee
The $\mu$-track events have two components: $N_{\mu\mu}$ from
$\mu$-tracks induced by muon neutrinos, and $N_{\mu\tau}$ from
$\mu$-tracks induced by tau neutrinos. The number of shower-like
events have three components:  $N_{sh_{had}}$ from hadronic showers
associated with NC interaction, $N_{sh_{em}}$ from electromagnetic
showers produced by CC interaction of $\nu_{e}$ and $N_{sh_{\tau}}$
from showers produced by  CC interaction of $\nu_{\tau}$ decaying
hadronically. So we can express $R$ as
\be
\text{R} = \frac{N_{\mu \mu} + N_{\mu \tau}}{N_{sh_{had}} + N_{sh_{em}} + N_{sh_{\tau}}}.
\label{eq:Rsum}
\ee
The $\mu$-tracks induced by $\nu_{\mu}(\bar{\nu}_{\mu})$  result from
the CC interaction of the neutrinos with the rock or the ice
underground.  The muons can travel a long distance before decaying;
the effective muon range $R_{\mu}$ depends on the initial energy
$E_{\mu}$ and the detection energy threshold $E_{\mu}^{th}$; in the
case of IceCube this threshold is $\sim\, 100$ GeV. The
$\mu$-track induced by $\nu_{\tau}(\bar{\nu}_{\tau})$ result from the
decay of a $\tau$ produced in a CC interaction into a $\mu$; this
decay has a  probability density $f(E_{\tau}, E_{\mu})$ and a
branching ratio $B = 17.8\%$. The expressions for $N_{\mu\mu}$ and
$N_{\mu\tau}$ are given by


\begin{equation}
N_{\mu \mu} =  \rho A N_{A} \int_{E_{\mu}^{th}}^{\ \infty} \int_{E_{\mu}^{th}}^{E_{\nu_{\mu}}} R_{\mu}(E_{\mu},E_{\mu}^{th}) P_{shad}(E_{\nu_{\mu}})\frac{dF_{\nu_{\mu}}}{dE_{\nu_{\mu}}} 
\frac{d\sigma^{CC}}{dE_{\mu}} dE_{\mu} dE_{\nu_{\mu}}\ \
 +\ \  [\nu_{\mu} \rightarrow \bar{\nu}_{\mu}].
\label{eq:muontrakcs}
\end{equation}

\bary
N_{\mu \tau} &=& B \rho A N_{A} \int_{E_{\mu}^{th}}^{\ \infty}
\int_{E_{\mu}^{th}}^{E_{\nu_{\tau}}}
\int_{E_{\mu}^{th}}^{\frac{E_{\tau}}{2}(1 + \beta)}R_{\mu}(E_{\mu},E_{\mu}^{th}) P_{shad}(E_{\nu_{\mu}})\nonumber\\
&&\times\frac{dF_{\nu_{\tau}}}{dE_{\nu_{\tau}}} 
\frac{d\sigma^{CC}}{dE_{\tau}} f(E_{\tau},E_{\mu}) dE_{\mu} dE_{\tau} dE_{\nu_{\tau}}\ \  + \ \ 
[\nu_{\tau} \rightarrow \bar{\nu}_{\tau}], 
\label{eq:muontracktau}
\eary

where the muon range is defined as
\begin{equation}
R_{\mu}(E_{\mu},E_{\mu}^{th}) = (2.6\ \text{Km}) \ln\left[\frac{2.0 + 4.2 \times 10^{-3} E_{\mu}}{2.0 + 4.2 \times 10^{-3} E_{\mu}^{th}}\right],
\label{eq:muonrange}
\end{equation}
and its probability density is given by
\begin{equation}
f(E_{\tau}, E_{\mu}) \simeq \frac{5}{3E_{\tau}} - \frac{3E_{\mu}^{2}}{E_{\tau}^{3}} + \frac{4E_{\mu}^{3}}{3E_{\tau}^{4}}\ .
\label{eq:probdens}
\end{equation}
The expression for $f(E_{\tau}, E_{\mu})$ is an approximation valid
for $\beta \rightarrow 1\ $ ($\gamma \gg 1$), 
where $\beta = \sqrt{1 - 1/\gamma^{2}\ } = \sqrt{1 - (\frac{m_{\tau}}{E_{\tau}})^{2}\ } $.
The number of shower-like events for the different kinds of processes are given by:

\begin{equation}
N_{sh_{had}} = \sum_{l=e,\mu,\tau} \rho A L N_{A} \left[ \int_{E_{sh}^{th}}^{\ \infty}P_{shad}(E_{\nu_{l}})  \frac{dF_{\nu_{l}}}{dE_{\nu_{l}}} \sigma^{NC} dE_{\nu_{l}}\  +\  
\int_{E_{sh}^{th}}^{\ \infty} P_{shad}(E_{\bar{\nu}_{l}}) \frac{dF_{\bar{\nu}_{l}}}{dE_{\bar{\nu}_{l}}} \sigma^{NC} dE_{\bar{\nu}_{l}} \right], 
\label{eq:sh_hadronic}
\end{equation}
\begin{equation}
N_{sh_{em}} = \rho A L N_{A} \left[ \int_{E_{sh}^{th}}^{\ \infty}P_{shad}(E_{\nu_{e}}) \frac{dF_{\nu_{e}}}{dE_{\nu_{e}}} \sigma^{CC} dE_{\nu_{e}}\ +\  
\int_{E_{sh}^{th}}^{\ \infty} P_{shad}(E_{\bar{\nu}_{e}})\frac{dF_{\bar{\nu}_{e}}}{dE_{\bar{\nu}_{e}}} \sigma^{CC} dE_{\bar{\nu}_{e}} \right],
\label{eq:sh_em}
\end{equation}
\begin{equation}
N_{sh_\tau} = (1-B)\rho A L N_{A} \left[ \int_{E_{sh}^{th}}^{\ \infty} P_{shad}(E_{\nu_{\tau}})\frac{dF_{\nu_{\tau}}}{dE_{\nu_{\tau}}} \sigma^{CC} dE_{\nu_{\tau}}\ +\  
\int_{E_{sh}^{th}}^{\ \infty}P_{shad}(E_{\bar{\nu}_{\tau}}) \frac{dF_{\bar{\nu}_{\tau}}}{dE_{\bar{\nu}_{\tau}}} \sigma^{CC} dE_{\bar{\nu}_{\tau}} \right],
\label{eq:sh_tau}
\end{equation}
where $\rho$ is the density of the detector medium, $A$ is the
effective area of the detector, $L$ is the length of the detector,
$N_{A}$ is the Avogadro's number and $dF_{\nu_{l}}/dE_{\nu_{l}}$ is
defined in Eq. (\ref{eq:Nu_Espectrum}). The normalization for this
equation, $N_{\nu_{l}}$, is proportional to the neutrino flux, for the
different flavors. Since  $dF_{\nu_{l}}/dE_{\nu_{l}}$ is evaluated in
the quotient of equation (\ref{eq:R}), the proportionality constant
cancels out. The total cross sections for $CC$ ($\sigma^{CC}$) and
$NC$  ($\sigma^{NC}$) shown in Figs. (\ref{fig:Xsec_neutrinos}) and
(\ref{fig:Xsec_antineutrinos}) are used to evaluate the $N_{sh_{had}}$
and $N_{sh_{em}}$.

In order to evaluate $d\sigma^{CC}/dE_{l}$ we performed an empirical
fit to the differential cross section presented in 
Fig. 4 of reference \cite{Bevan.:2007pt}, which is given as:
\begin{equation}
 \frac{1}{\sigma^{CC}}\frac{d\sigma^{CC}}{dy} = N_0  \left\{ \begin{array}{lcc}
 b_1\ y^{-a_1} & \ \ \ \ \text{if} & \  y < y_{cut}\\ \\
 b_2\ y^{-a_2} & \ \ \ \ \text{if} & \  y \geq y_{cut}\ ,
 \end{array}
 \right .
 \label{eq:dsdy}
\end{equation}
where $N_0$ is the normalization,
\begin{equation}
 y = \frac{E_{\nu_{l}}-E_{l}}{E_{\nu_{l}}},
\label{eq:y}
\end{equation}
and
\begin{equation}
y_{cut} = \exp \left( \frac{\log b_1 - \log b_2}{a_1 - a_2} \right).
\label{eq:ycut}
\end{equation}
The parameters in Eq.(\ref{eq:dsdy}) are as follows:
\begin{equation}
a_1 =  -0.0163\ x^2 + 0.3877\ x - 1.1905 ,
\label{eq:a1}
\end{equation}
\begin{equation}
a_2 = -0.0222\ x^2 + 0.4222\ x - 0.9833,
\label{eq:a2}
\end{equation}
\begin{equation}
b_1 =  0.0168\ x^2 - 0.3683\ x + 2.0038 ,
\label{eq:b1}
\end{equation}
\begin{equation}
b_2 = 0.0139\ x^2 - 0.2739\ x + 1.4233,
\label{eq:b2}
\end{equation}
and 
\begin{equation}
x = \log_{10}(E_{\nu_l}/\text{GeV}).
\label{eq:x}
\end{equation}
The normalization is set such that 
\begin{equation}
\int_{0}^{1} \left( \frac{1}{\sigma^{CC}}\frac{d\sigma^{CC}}{dy} \right) dy = 1\ .
\label{eq:Norm}
\end{equation}
We compare our fit with the data presented in 
reference\cite{Bevan.:2007pt} which are shown in Fig. \ref{fig:dsdy}.

After performing the necessary change of variable from $E_{l}$ to $y$,
one can evaluated the integrals numerically. The neutrino-flavor
ratios, $R$, obtained after propagating the neutrinos from the source, all
the way up to the detector, for different combinations of the
parameters involved, and for different energies are used as input for
the calculation.

\begin{figure}
\begin{center}
\includegraphics[width= 80mm, height=80mm]{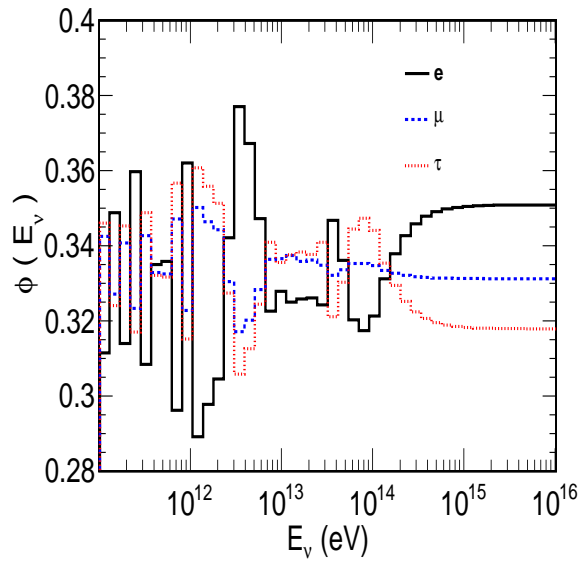}   
\caption{Neutrino flux $\Phi_{\nu} (E_{\nu})$ as a function of neutrino energy $E_{\nu}$ in the detector.}
\label{fig:avgnuflux}
\end{center}
\end{figure}
\begin{figure}
\begin{center}
\includegraphics[width= 80mm, height=80mm]{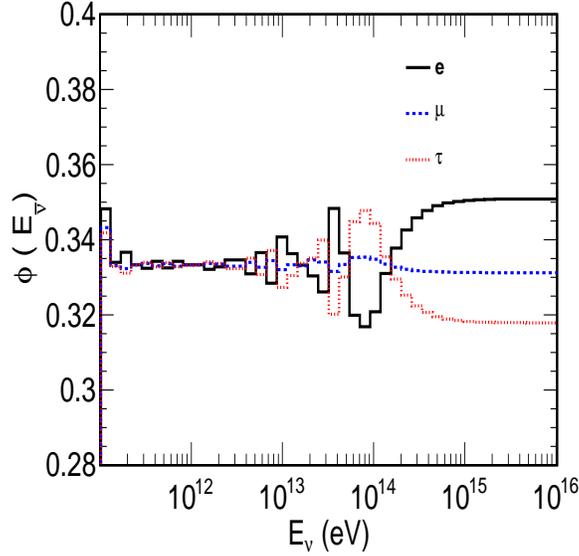}   
\caption{Antineutrino flux $\Phi (E_{\bar \nu})$ as a function of neutrino energy $E_{\nu}$ in the detector.}
\label{fig:avgantinuflux}
\end{center}
\end{figure}

\section{Results}
\label{sec:4}

\begin{figure}
\begin{center}
\includegraphics[width= 80mm, height=80mm]{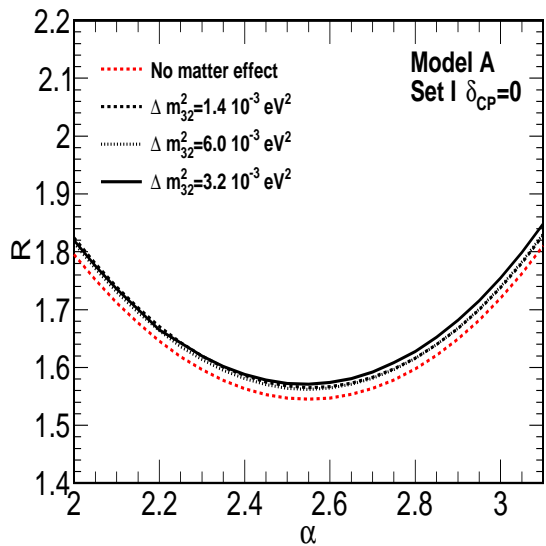}   
\caption{The track-to-shower ratio $R$ as a function of the spectral
  index $\alpha$ for $\delta_{CP}=0$ in Model-A.}
\label{fig:R_SetI_dCP0}
\end{center}
\end{figure}

\begin{figure}
\begin{center}
\includegraphics[width= 80mm, height=80mm]{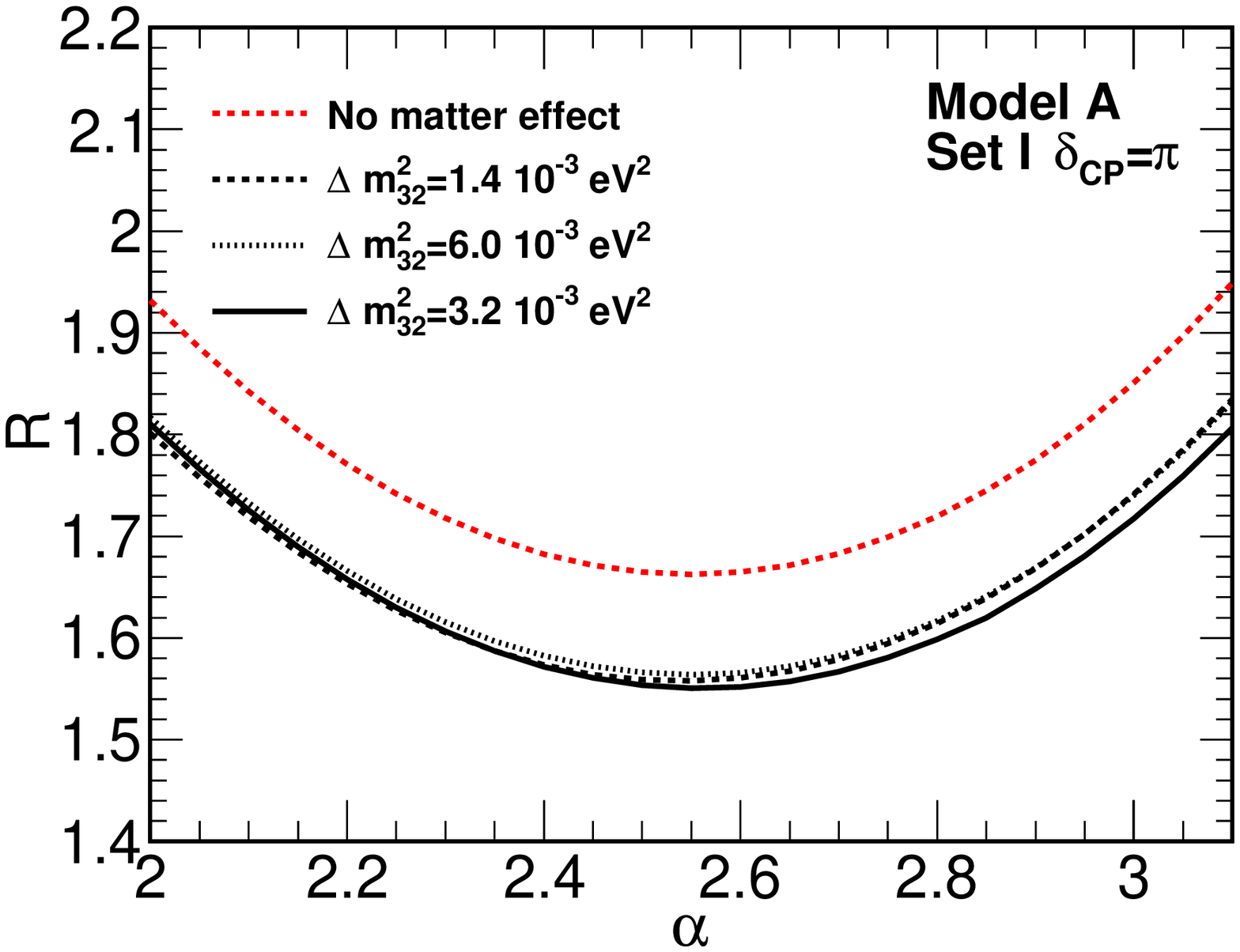}   
\caption{The track-to-shower ratio $R$ as a function of the spectral
  index $\alpha$ for $\delta_{CP}=\pi$ in Model A. }
\label{fig:R_SetI_dCPpi}
\end{center}
\end{figure}

\begin{figure}
\begin{center}
\includegraphics[width= 80mm, height=80mm]{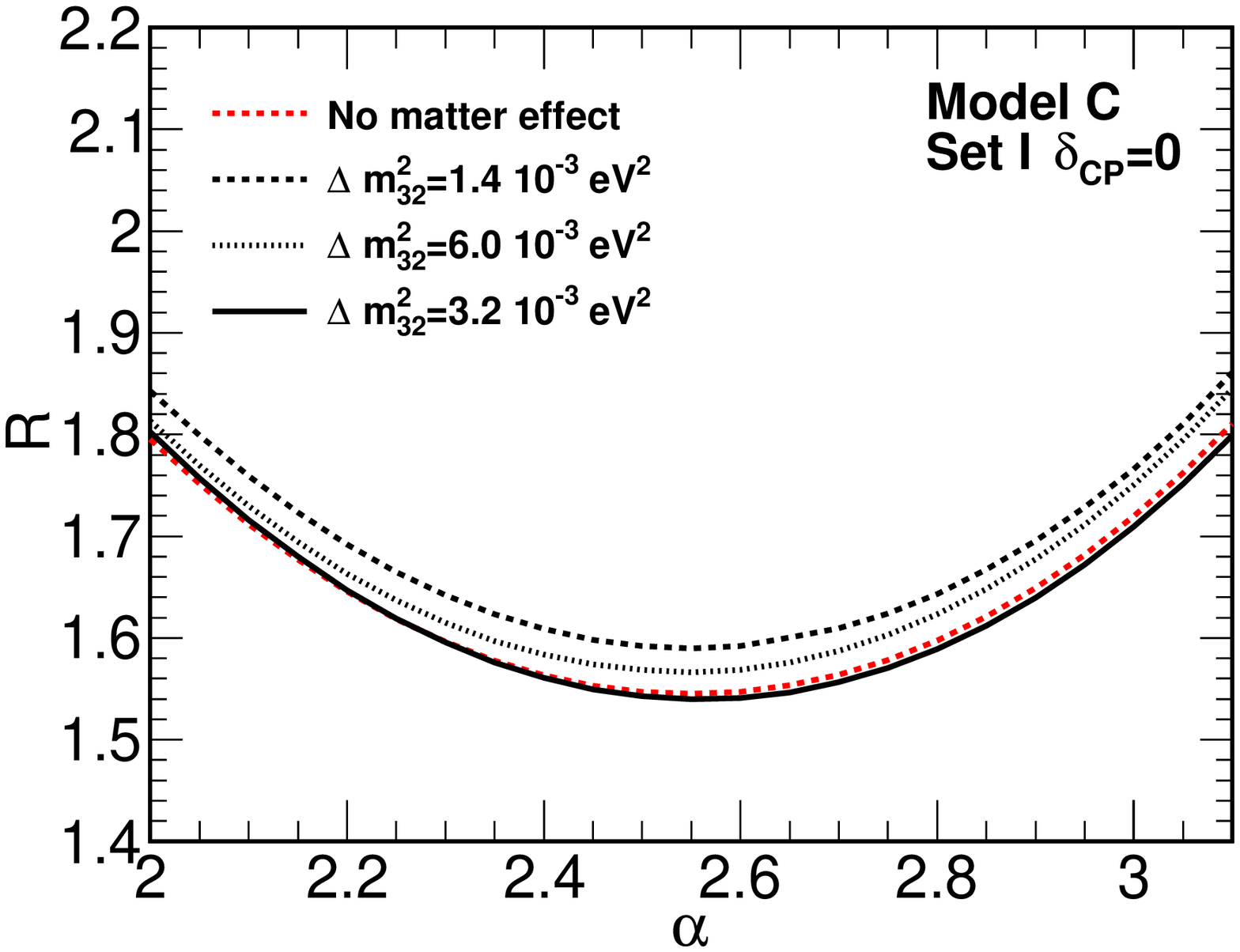}   
\caption{Same as Fig. \ref{fig:R_SetI_dCP0} for model C.}
\label{fig:R_SetII_dCP0}
\end{center}
\end{figure}

\begin{figure}
\begin{center}
\includegraphics[width= 80mm, height=80mm]{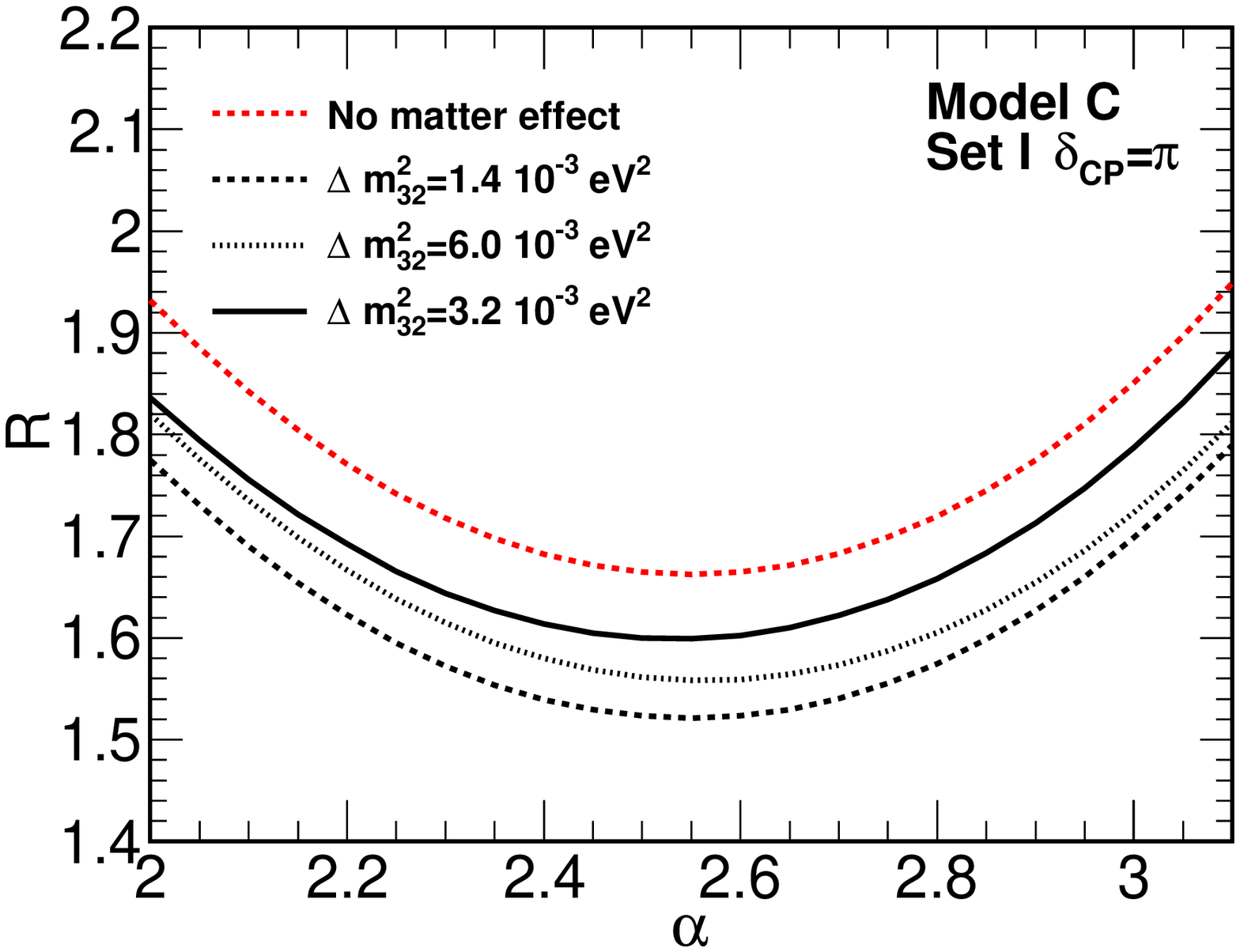}   
\caption{Same as Fig. \ref{fig:R_SetI_dCPpi} for model C. }
\label{fig:R_SetII_dCPpi}
\end{center}
\end{figure}

\begin{figure}
\begin{center}
\includegraphics[width= 80mm, height=80mm]{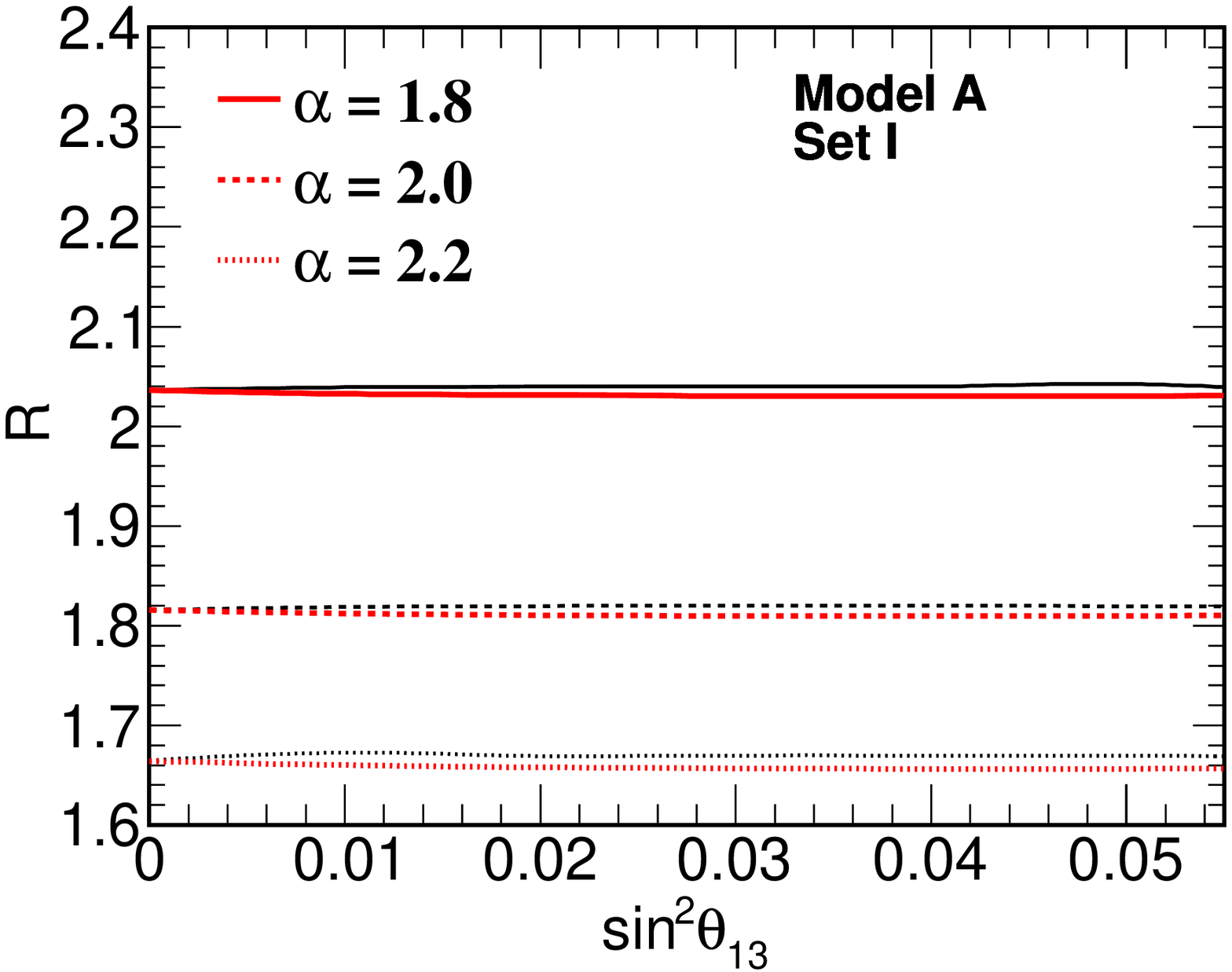} 
\caption{The track-to-shower ratio $R$ as a function of $sin^2\theta_{13}$
  in Model-A for the parameter Set-I with $\Delta m^2_{32}=3.2\times
  10^{-3}\, eV^2$. The black curve is for
  $\delta_{CP}=0$ and red one is for $\delta_{CP}=\pi$.}

\label{fig:RvsSin2th13_ModelA_SetI}
\end{center}
\end{figure}

\begin{figure}
\begin{center}
\includegraphics[width= 80mm, height=80mm]{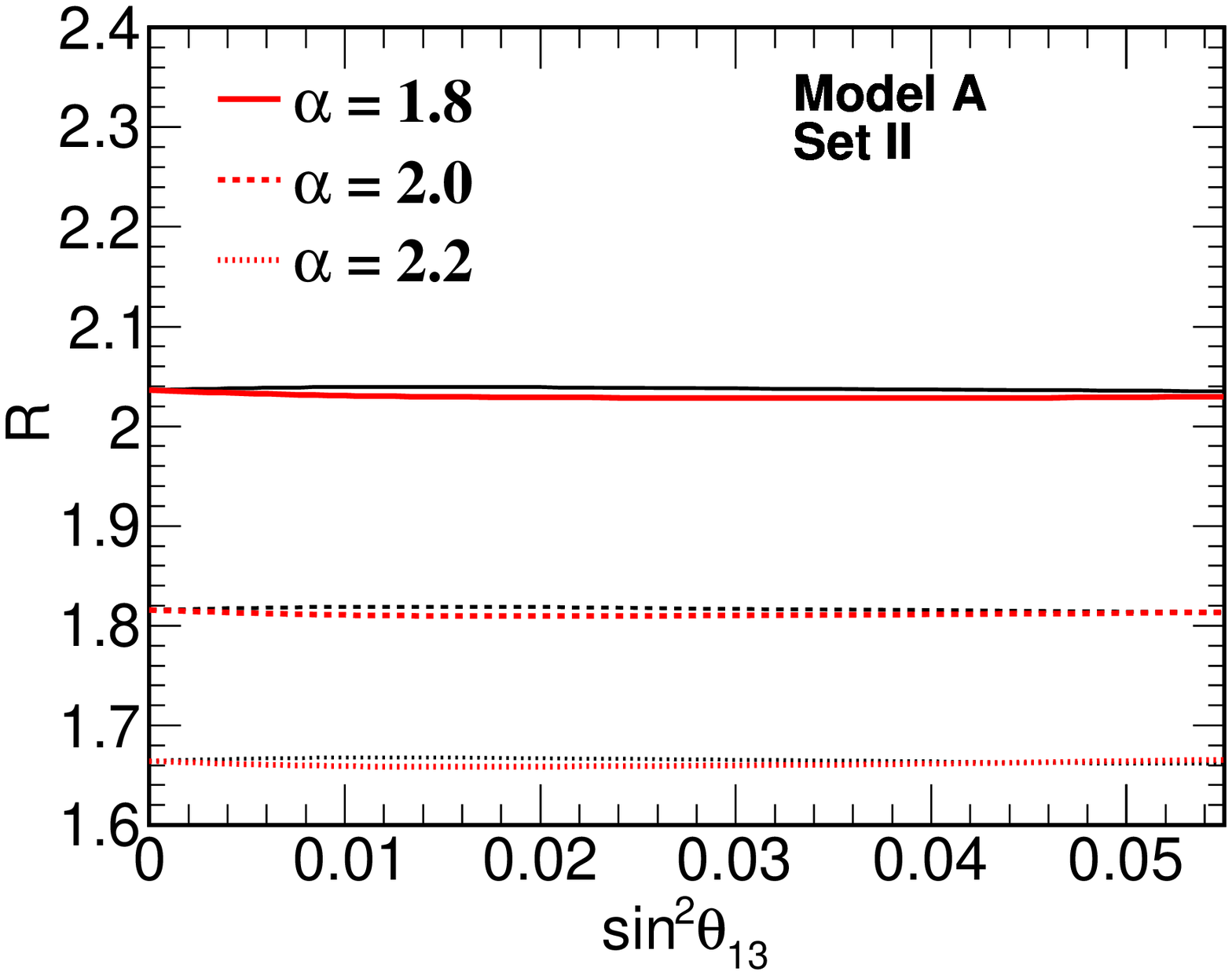}   
\caption{The track-to-shower ratio $R$ as a function of $sin^2\theta_{13}$
  in Model-A for the parameter Set-II with $\Delta m^2_{32}=3.2\times
  10^{-3}\, eV^2$. The black curve is for
  $\delta_{CP}=0$ and red one is for $\delta_{CP}=\pi$.}
\label{fig:RvsSin2th13_ModelA_SetII}
\end{center}
\end{figure}

\begin{figure}
\begin{center}
\includegraphics[width= 80mm, height=80mm]{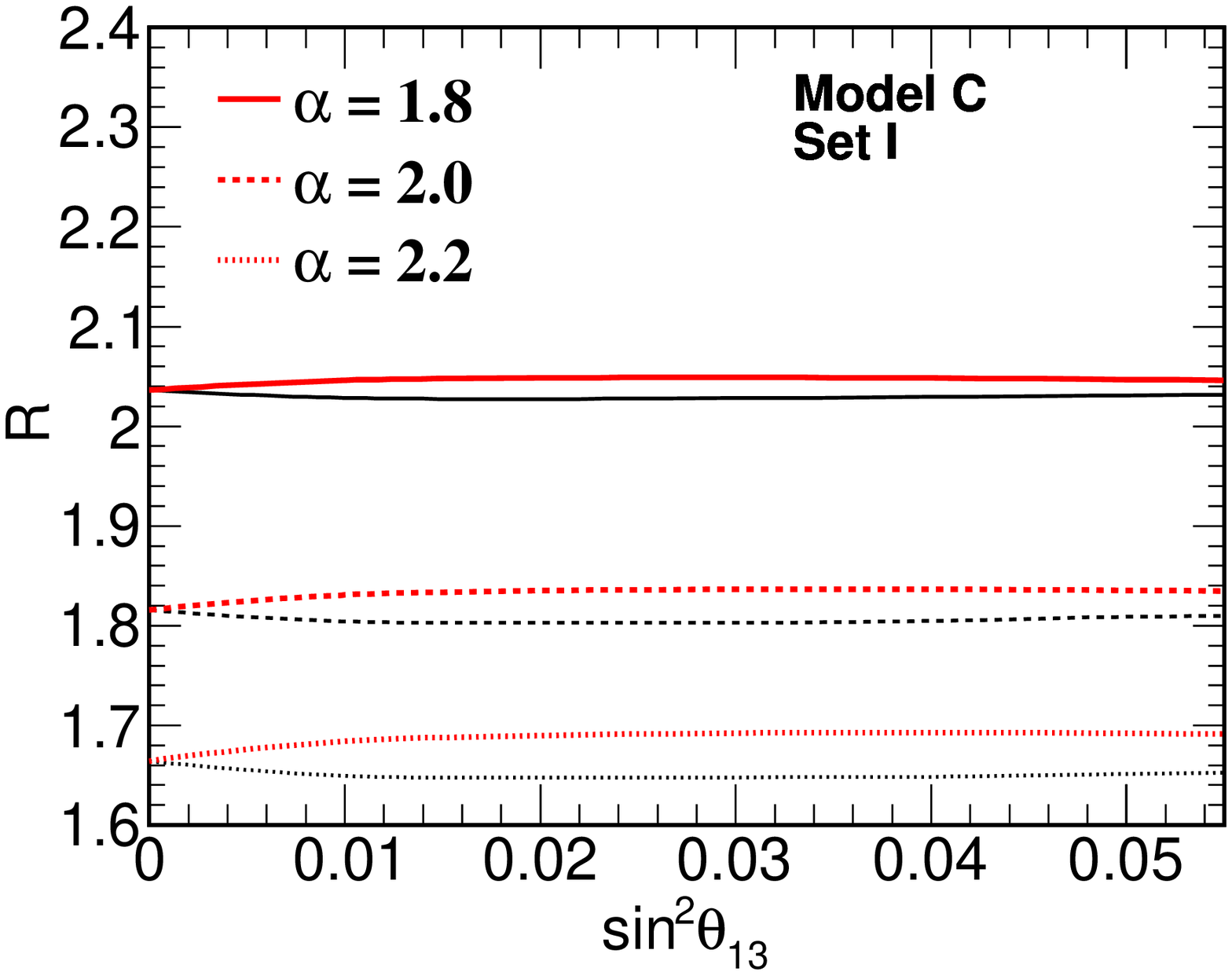} 
\caption{The track-to-shower ratio $R$ as a function of $sin^2\theta_{13}$
  in Model-C for the parameter Set-I with $\Delta m^2_{32}=3.2\times
  10^{-3}\, eV^2$. The black curve is for
  $\delta_{CP}=0$ and red one is for $\delta_{CP}=\pi$.}
\label{fig:RvsSin2th13_ModelC_SetI}
\end{center}
\end{figure}

\begin{figure}
\begin{center}
\includegraphics[width= 80mm, height=80mm]{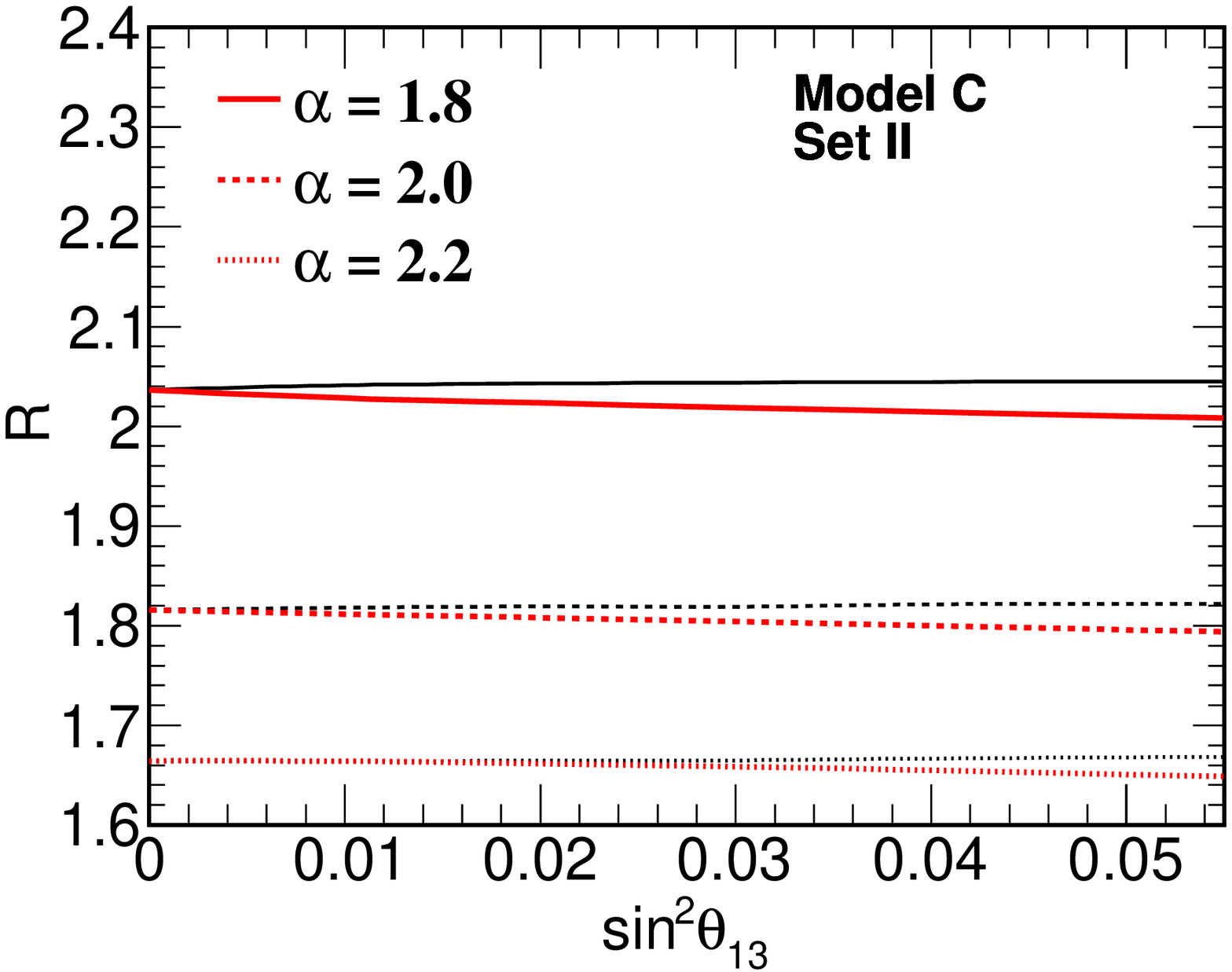}   
\caption{The track-to-shower ratio $R$ as a function of $sin^2\theta_{13}$
  in Model-C for the parameter Set-II with $\Delta m^2_{32}=3.2\times
  10^{-3}\, eV^2$. The black curve is for
  $\delta_{CP}=0$ and red one is for $\delta_{CP}=\pi$.}
\label{fig:RvsSin2th13_ModelC_SetII}
\end{center}
\end{figure}

\begin{figure}
\begin{center}
\includegraphics[width= 80mm, height=80mm]{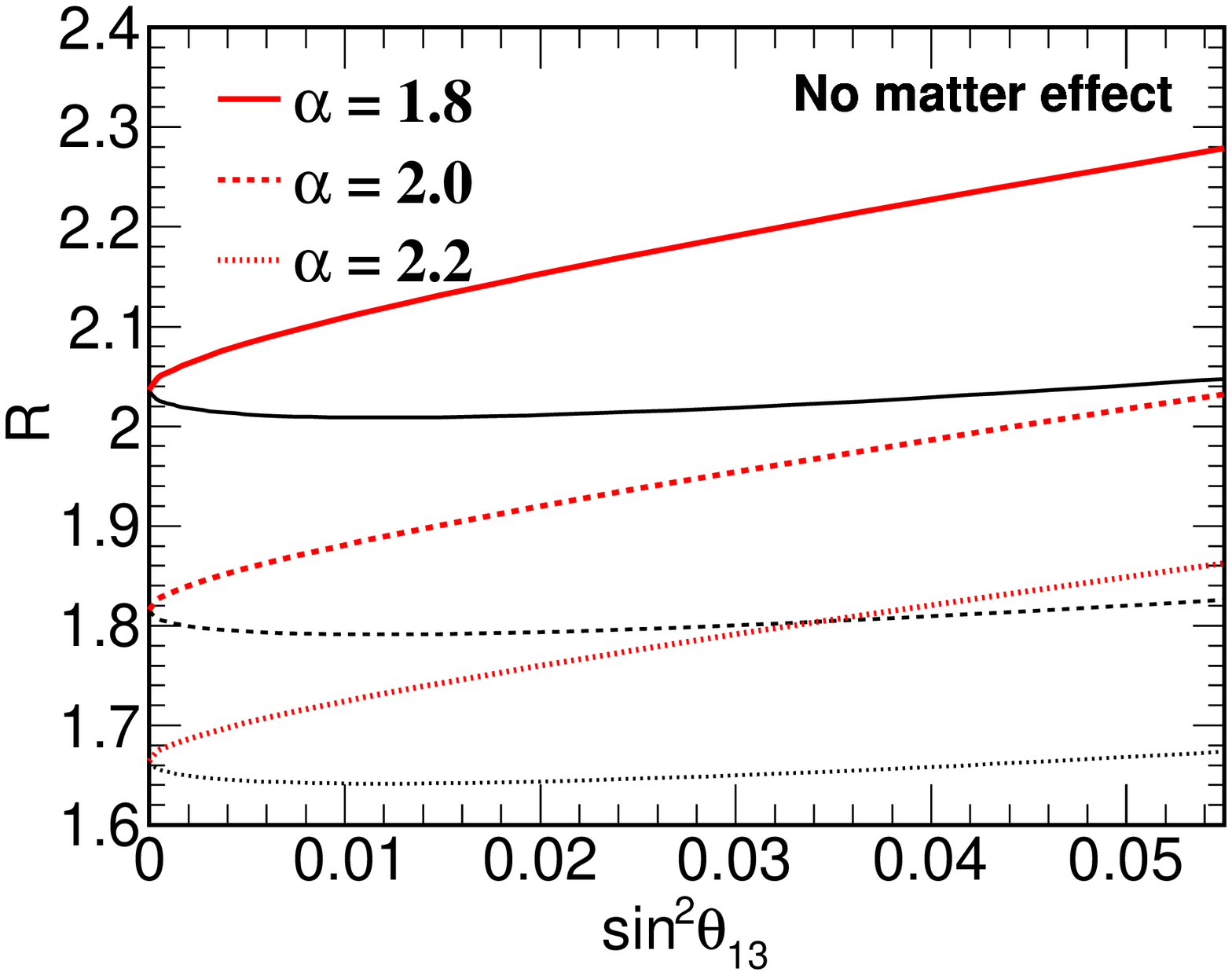} 
\caption{The track-to-shower ratio $R$ as a function of
  $sin^2\theta_{13}$ with no matter effect. Hre also the black curve is for
  $\delta_{CP}=0$ and red one is for $\delta_{CP}=\pi$. We take $\Delta m^2_{32}=3.2\times
  10^{-3}\, eV^2$.}
\label{fig:RvsSin2th13_nomatter}
\end{center}
\end{figure}

As can be seen from Figs.\ref{fig:Neut_FluX_S1}
and \ref{fig:Neut_FluX_S2},  the normalized flux of
neutrinos and antineutrinos in the detector depends on energy. For
the calculation of the ratio $R$ we need the neutrino flux
$dF_{\nu}/dE_{\nu}$. Neither we  know the exact form of it nor the
spectral index $\alpha$. But by considering the neutrino flux ratio
1:2:0 at the source, then propagating these neutrinos through the
presupernova matter we calculated the normalized flux on the surface
of the star in Paper-I. Here, we take these normalized flux and
propagate the neutrinos through the distance between the source and
the Earth, where Earth's matter effect is included and calculate the
normalized flux of these neutrinos and antineutrinos in the
detector. For the calculation of the  track-to-shower ratio $R$ of
Eq.(\ref{eq:R}) we use these fluxes. But instead of calculating the
flux for each energy, we divide the whole energy range to energy bins
as $\Delta E_{\nu}=0.3 E_{\nu}$ i.e. \%30 energy resolution. Within
each bin the flux is constant which we take by averaging the flux
in the same energy bin. Here we have shown these avarage neutrino and
antineutrino fluxes in  
Figs.\ref{fig:avgnuflux} and \ref{fig:avgantinuflux}. From these
figures, it is observed that the average neutrino and antineutrino
fluxes are different for $E_{\nu} < 2\times 10^{13}$ eV. Finally, we consider two values of
the CP violating phase $\delta_{CP}=0$ and $\pi$ to see the change in
$R$. The upper limit of the $E_{\nu}$ is taken to be $10$ PeV to
evaluate the neutrino energy integrals. The following values are
considered for the IceCube detector in our calculation:
density of ice $\rho=0.051\, g\,cm^{-3}$, detector area
$A=10^{10}\,cm^2$ and the detector length $L=10^5\, cm$.
The results are
presented in Figs. \ref{fig:R_SetI_dCP0}  to \ref{fig:RvsSin2th13_nomatter}.




In Figs. \ref{fig:R_SetI_dCP0}  to \ref{fig:R_SetII_dCPpi},  we have shown $R$ as a function of the spectral index
$\alpha$ for models A and C. In these figures we also include {\it no
  matter effect} which implies: at the source we consider the flux
ratio 1:2:0 and these neutrinos propagate up to the detector
in vacuum. For our convenience we define the track-to-shower ratio for no matter effect as $R_0$. For $\delta_{CP}=0$ we
found that $R_0 \le  R$ for any given
value of $\alpha$. Also the gap between $R$ and $R_0$ is small. On the
other hand, for $\delta_{CP}=\pi$, always we found  $R_0
> R$ and the gap is bigger. 
The value of $R$ is minimum around $\alpha=2.6$  which is
independent of whether we consider matter effect or not. 
We have also shown for three different
$\Delta m^2_{32}$ values, which shows that there is very little
variation in $R$. This minimum
value of $R$  is also independent  of $\Delta m^2_{32}$. 
The order in which $R$ is arranged for different
$\Delta m^2_{32}$ values  reverses by going from $\delta_{CP}=0$ to
$\pi$, which can be seen by comparing Fig. \ref{fig:R_SetI_dCP0} with
Fig. \ref{fig:R_SetI_dCPpi}  in model A and similarly
Fig. \ref{fig:R_SetII_dCP0}  with Fig. \ref{fig:R_SetII_dCPpi} in
model C. Here we have omitted the results from model B because the results are very
similar to model A.

In Figs. \ref{fig:RvsSin2th13_ModelA_SetI} to
\ref{fig:RvsSin2th13_ModelC_SetII} we have shown the variation of $R$
as a function of $sin^2{\theta_{13}}$ in models A and C for three
different values of the spectral index $\alpha$. In these plots we
observe that the ratio $R$ is almost constant for a given $\alpha$ and
for both $\delta_{CP}=0$ and $\pi$, as we vary
$sin^2{\theta_{13}}$ for all the models. Also the value of $R$ is
higher for smaller $\alpha$.

We have also shown the $R$ as a function of $sin^2{\theta_{13}}$  for
no matter effect in Fig.\ref{fig:RvsSin2th13_nomatter}. This shows a
clear difference between $\delta_{CP}=0$ (lower curve)  and
$\delta_{CP}=\pi$ (upper curve) for each $\alpha$. These two curves diverge from the
point $\theta_{13}=0$ as can be seen from the plots in
Fig.\ref{fig:RvsSin2th13_nomatter}. Comparison of the matter effect
(from Figs. \ref{fig:RvsSin2th13_ModelA_SetI}  to
\ref{fig:RvsSin2th13_ModelC_SetII}) with the no matter effect
Fig.\ref{fig:RvsSin2th13_nomatter} shows that the $\delta_{CP}=\pi$
contribution is very much suppressed in matter compared to
$\delta_{CP}=0$ contribution and makes them almost the same. This
shows that the track-to-shower ratio $R$ for high energy neutrinos in
IceCube is probably almost blind to CP violating phases when Earth matter
effect is taken into account.

\section{Summary}
\label{sec:4}

A very small fraction ($\le\, 10^{-3}$) of the core collapse supernovae
can produce GRBs by launching a successful jet. Although the majority of these
core collapse can not produce GRBs, very high energy neutrinos can
easily be produced in their choked jets. These neutrinos propagating
through the over burden matter can undergo oscillation and the flux
ratio on the surface of the star can be different from the point where
these neutrinos were produced. The Mpc long baseline, from the surface of the star to the
surface of the Earth, these neutrinos will have vacuum
oscillation. Before reaching to the detector from the opposite side of
the Earth, these neutrinos will cross the dimeter of the Earth and can
undergo again matter oscillation. By considering the realistic density
profile of the Earth we have extended our previous work to study
numerically the three neutrino oscillation and evaluate the change in the flux ratio
in the detector. Depending on the energy of these
neutrinos, there can also be shadowing effect and neutrinos above few
PeV can be completely absorbed. 
In this work we have done a through analysis of the high energy
neutrino propagation in the Earth before reaching to the detector by
taking into account the shadowing effect. The track-to-shower ratio
$R$ is calculated for these high energy neutrinos. In the calculation
of $R$ we have included the shadowing effect and the contribution of
muon track produced by the high energy $\tau$ lepton decay around the
IceCube detector. These $\tau$ leptons are produced due to the CC
interaction of $\nu_{\tau}$ with the surround rock and ice of the
detector. We have studied the variation of $R$ when the
spectral index $\alpha$ and the mixing angle $sin^2\theta_{13}$
vary. We found that $R$ has a minimum around $\alpha=2.6$ and is
independent of whether we consider matter effect or not. This minimum
value of $R$  is also independent  of $\Delta m^2_{32}$ value. We
observed that the ratio $R$ is different for $\delta_{CP}=0$ and
$\pi$ when no matter effect is considered. But when Earth matter
contribution is taken into account, the $R$ value is 
almost blind to these different CP phases.

S.S. is thankful to Departamento de Fisica de Universidad de los
Andes, Bogota, Colombia, for their kind hospitality during his several
visits. This work is partially supported by DGAPA-UNAM (Mexico) Project 
No. IN103812.

\end{document}